\documentclass[prb,aps,eqsecnum]{revtex4}% Physical Review B
\usepackage{graphicx}
\usepackage{epsf}
\newcommand{\e}{{\rm e}}
\newcommand{\ep}{\varepsilon}

\newcommand{\bea}{\begin{eqnarray}}
\newcommand{\eea}{\end{eqnarray}}
\newcommand{\be}{\begin{equation}}
\newcommand{\ee}{\end{equation}}
\newcommand{\ba}{\begin{eqnarray}}
\newcommand{\ea}{\end{eqnarray}}
\newcommand{\nn}{\nonumber}
\newcommand{\la}{\label}
\newcommand{\lb}{\label}

\newcommand{\hT}{\hat{T}}
\newcommand{\hV}{\hat{V}}
\newcommand{\hU}{\hat{U}}

\newcommand{\htH}{\hat{H}_A}
\newcommand{\ld}{{\lambda}}
\newcommand{\dt}{{\Delta\theta}}
\newcommand{\pa}{\partial}
\newcommand{\bA}{{\bf A}}
\newcommand{\br}{{\bf r}}
\newcommand{\bp}{{\bf p}}
\newcommand{\bL}{{\bf L}}
\newcommand{\bO}{{\bf \Omega}}

\def\t1{e_T}
\def\v1{e_V}

\def\ct{e_{TTV}}
\def\cv{e_{VTV}}
\def\tt{e_{TTTTV}}
\def\tv{e_{VTTTV}}
\def\vt{e_{TTVTV}}
\def\vv{e_{VTVTV}}
\def\cT{{\cal T}}

\def\fft#1#2{{#1 \over #2}}

\def\tH{{H_A}}

\begin{document}
%\tightenlines
\title{The physics of symplectic integrators: perihelion advances and
symplectic corrector algorithms}

\author{Siu A. Chin}

\affiliation{Department of Physics, Texas A\&M University,
College Station, TX 77843, USA}

%\date{\today}
\begin{abstract}

Symplectic integrators evolve dynamical systems according to modified
Hamiltonians whose error terms are also well-defined Hamiltonians.
The error of the algorithm is the sum of each error 
Hamiltonian's perturbation on the exact solution.
When symplectic integrators are applied to the Kepler problem,
these error terms cause the orbit to precess.
In this work, by developing a general method of computing the
perihelion advance via the Laplace-Runge-Lenz vector even for
non-separable Hamiltonians, I show that the precession error
in symplectic integrators can be computed analytically.
It is found that at each order, each paired error Hamiltonians
cause the orbit to precess oppositely by exactly the same amount
after each period. Hence, symplectic corrector, or process integrators,
which have equal coefficients for these paired error terms,
will have their precession errors exactly cancel after each period. 
Thus the physics of symplectic integrators determines the optimal
algorithm for integrating long time periodic motions.

\end{abstract}
\maketitle
\section {Introduction}

Numerical methods for solving physical problems are generally not
expected to contain interesting physics. They are viewed as mere
means, or recipes, of arriving at a needed numerical solution. This
is because most numerical methods are based on matching
Taylor series, whose error terms have little to do with physics. 
By contrast, symplectic integrators
solve dynamical problems by approximating the original Hamiltonian by a
modified Hamiltonian whose error terms are also well-defined
Hamiltonians. In the past, these error terms are just formal
entities destine to be eliminated by order-conditions, and are
rarely studied in their own right. Here, we show that a
comprehensive study of the error Hamiltonians in the Kepler problem
gives insights into the working of symplectic integrators 
and makes manifest, ways of optimizing them.

Symplectic Integrators (SI) \cite{yoshi,mcl02,hairer} despite their
excellent conservation properties, are not immune from the
fundamental phase error when solving the Kepler problem. While the
energy error is periodic, the phase error can accumulate and grow
linearly with time\cite{shita,gladman,cano}. One manifestation of 
the phase error is the ``perihelion advance" of the numerically 
computed elliptical orbit. This error is particularly pernicious 
when contemplating long time integration of periodic motions.
No matter how small the initial time step, 
the orbital precession error can accumulate after each period 
and grow linearly without bound.

In the Kepler problem, the energy error causes the
length of Laplace-Runge-Lenz (LRL) vector to oscillate and
the phase error causes the vector to rotate\cite{chinkid}.
While the energy error has been studied extensively,
little is known about the phase error and its cause.
This is reflected in the historical development of
symplectic integrators; most early integrators are not well-tuned for
the reduction of phase errors. For example, when solving the Kepler
problem, the first fourth-order, Forest-Ruth\cite{forest} algorithm has
a much larger precession error per period than the standard 
fourth-order Runge-Kutta algorithm\cite{chinkid}. Even the much 
improved McLachan integrator\cite{mcl95} has a larger precession 
error than that of Runge-Kutta\cite{chinsante}.

In this work, we present a detailed study of the precession error 
due to each error Hamiltonian (up to fourth order) 
on Kepler's orbit. Based on Sivardi\`ere's 
method\cite{sivar} of computing the rotation of the LRL vector, 
we develop a comprehensive treatment
of perihelion advance due to any perturbing Hamiltonian, including
non-separable ones. We show analytically that paired error terms of
the form $\{T,Q\}$ and $\{V,Q\}$ rotate the LRL vector oppositely by
exactly the same amount after each period. Here $T$ and $V$ are the
kinetic and potential energy functions of the Kepler Hamiltonian,
$\{A,B\}$'s are Poisson brackets, and $Q$'s are higher order
Poisson brackets of $T$ and $V$. Algorithms with equal
coefficients for these paired error terms would therefore have their
precession errors precisely cancel after each period.
This class of algorithm has been previously identified\cite{chincor}
as symplectic corrector\cite{wis96,mcl962}, 
or process\cite{mar96,mar97,blan99} algorithms. Symplectic
corrector algorithms were originally derived for their 
computational efficiency; this work further identify 
them as a class of integrators with
periodic precession errors. Thus the physical effects of these error 
Hamiltonians provide the needed insight for devising 
optimal integrators with periodic energy and phase conservation. 

For the Kepler problem, highly specialized algorithms\cite{ymin02,ymin04}
can be devised to exactly conserve energy and the rotation of the LRL vector. 
However, these algorithms do not limit the growth of the phase error in time. 
At a given time, the particle is at the wrong point of the trajectory, 
despite the fact it is constrained to move on the correct trajectory. Also, 
the phase errors in these algorithms are only second order in 
$\Delta t$ and cannot be systematically improved to higher orders. This
work solves the Kepler problem to fourth order in both the energy 
and the precession error and illustrates a general philosophy of 
allowing the physics of the problem to dictate the type of 
algorithm to be used for its solution.  

In the next section, we will summarize needed results on the error
structure of symplectic integrators. This is followed by Section III
where we derive analytical expressions for the rotation
angle of the LRL vector per period due to error Hamiltonians up to the fourth
order. In this work, we systematize and generalize Sivadi\`ere's 
method\cite{sivar} of computing orbital precession to include any 
angular-momentum conserving Hamiltonians, even non-separable ones.
In Section IV, we numerical verify these theoretical predictions. 
In Section V, we derive second and fourth-order corrector algorithms 
with demonstrated periodic precession errors. Some conclusions and 
directions for future research are given in Section VI.

\section {Error Hamiltonians of symplectic integrators }

    Symplectic integrators for evolving the standard Hamiltonian
%%%%%
\be
H({\bf q},{\bf p}) = T({\bf p})+V({\bf q}),\qquad {\rm with}\qquad
T({\bf p}) = \fft{1}{2}p_i p_i\, ,\la{ham}
\ee
%%%%%
can be derived\cite{yoshi} by approximating the system's short-time
evolution operator via a product of elemental evolution
operators  ${\rm e}^{\,\ep\, \hT}$ and ${\rm e}^{\,\ep\, \hV}$ 
via
%%%%%
\be
{\rm e}^{\ep(\hT+\hV)}\approx \prod_{i=1}^N
{\rm e}^{t_i\ep\hT}{\rm e}^{v_i\ep\hV}, \la{prod}
\ee
%%%%%
where each Lie operator\cite{dragt} $\hat Q$ associated with variable $Q$
acting on any other dynamical variable $W$ is defined by the Poisson
bracket \be \hat{Q}\, W=\{W,Q\}\, . \la{liop} \ee For a given set of
factorization coefficients $\{t_i,v_i\}$, the product on the RHS of
(\ref{prod}) then produces a ordered sequence of displacements
%%%%%
\ba
&&p_i(\ep)={\rm e}^{\,\ep\, \hV}p_i=p_i-\ep\,\fft{\pa V}{\pa q_i} ,\nn\\
&&q_i(\ep)={\rm e}^{\,\ep\, \hT}q_i= q_i+\ep\,\fft{\pa T}{\pa p_i},
\la{pqsh}
\ea
which defines the resulting algorithm. For a more detailed description,
see Ref.\cite{yoshi} and Ref.\cite{chinsante}. For the study of
time-reversible Hamiltonians, we will only consider 
time-reversible, symmetric factorization schemes
such that  either $t_1=0$ and
$v_i=v_{N-i+1}$, $t_{i+1}=t_{N-i+1}$, or $v_N=0$ and
$v_i=v_{N-i}$, $t_{i}=t_{N-i+1}$. (The use of asymmetric schemes
to study time-reversible Hamiltonians may introduce unphysical 
and unnecessary distortion\cite{chinsanpl} of the phase space 
at fintie $\Delta t$.) 

The product of operators in (\ref{prod}) can be combined by use of the
Baker-Campbell-Hausdorff (BCH) formula to give
\be
\prod_{i=1}^N
{\rm e}^{t_i\ep \hT}{\rm e}^{v_i\ep \hV}
={\rm e}^{\ep\htH},
\la{tbopt}
\ee
where $\htH$ is Hamiltonian operator
of the algorithm.
By the repeated use of (\ref{liop}), one can deduce
the Hamiltonian function $H_A$ corresponding to the
Lie operator $\htH$:
\bea
\tH &=& e_{T} T+e_{V} V\,
+\,\ep^2\,\left(\,\ct\,\{T^2\,V\}+\cv\,\{V\,T\,V\}\,\right)\,\nn \\
&&+\,\ep^4\,\left(\,\tt\,\{T\,T^3\,V\}
+ \tv\,\{V\,T^3\,V\}\right.\, \nn \\
&&\quad\quad\quad\left. +\,\vt\,\{T\,(T\,V)^2\}\,
+\,\vv\,\{V\,(T\,V)^2\}\,\right)\,+\,\dots\, ,
\la{hft4th}
\eea
where $\{TTV\}=\{T,\{T,V\}\}$, $\{T(TV)^2\}=\{T,\{T,\{V,\{T,V\}\}\}\}$ etc.,
are condensed Poisson bracket notations. This is the Hamiltonian function 
conserved by the algorithm.
The error coefficients
$e_{T}$, $e_{V}$, $\vv$, {\it etc}., are algorithm specific,
calculable from knowing	the
$\{t_i,v_i\}$ coefficients\cite{nosix}. In particular,
\be
e_{T}=\sum_{i=1}^Nt_i,\quad e_{V}=\sum_{i=1}^N v_i.
\ee
Thus all algorithms must have $e_T=1=e_V$ in order to reproduce the original
Hamiltonian. This will always be assumed.
The Poisson brackets reflect properties of the original
Hamiltonian:\cite{chinsante}
\bea
\{TTV\}\}&=& p_{i}V_{ij}p_{j}\, ,\nn\\
\{VTV\}\}&=&-V_{i}V_{i}\, ,\nn\\
\{T\,(T\,V)^2\}&=&-2p_i(V_{ikj}V_k+V_{ik}V_{kj})p_j\, ,\nn\\
\{V\,(T\,V)^2\}&=&2 V_iV_{ij}V_j\, ,\nn\\
\{T\,T^3\,V\}&=&p_ip_jp_kp_lV_{ijkl}\, ,\nn\\
\{V\,T^3\,V\}&=&-3p_ip_jV_{ijk}V_k\, .
\la{errsix}
\eea
To emphasize that these error terms are Hamiltonians, we will also
denote $H_{TTV}=\{T,\{T,V\}\}$, $H_{TTTTV}=\{T\,T^3\,V\}$,
etc..

For a central potential
\be
V({\bf q})=V(r),
\ee
one can easily verify that
\ba
V_{i}&=&V^\prime\hat\br_i,\nn\\
V_{ij}&=&U\delta_{ij}
+(V^{\prime\prime}-U)\hat\br_i\hat\br_j,
\la{vij}\\
V_{ijk}&=&U^\prime
(\delta_{ij}\hat\br_k+\delta_{jk}\hat\br_i+\delta_{ki}\hat\br_j)
+( V^{\prime\prime\prime}-3U^\prime) \hat\br_i\hat\br_j\hat\br_k,
\la{vijk}\\
V_{ijkl}&=&r^{-1}U^\prime(\delta_{ij}\delta_{kl}+\delta_{jk}\delta_{il}+\delta_{ki}\delta_{jl} )
+(V^{\prime\prime\prime\prime}-6U^{\prime\prime}+3r^{-1}U^\prime)
\hat\br_i\hat\br_j\hat\br_k\hat\br_l\nn\\
&+&(U^{\prime\prime}-r^{-1}U^\prime)(\delta_{ij}\hat\br_k\hat\br_l
         +\delta_{jk}\hat\br_i\hat\br_l+\delta_{ki}\hat\br_i\hat\br_l
+\delta_{il}\hat\br_j\hat\br_k+\delta_{jl}\hat\br_k\hat\br_i+
\delta_{kl}\hat\br_i\hat\br_j
         )
\la{vijkl}
\ea
where we have defined
\be
U(r)=\frac{V^\prime(r)}r.
\ee
The forms (\ref{vij})-(\ref{vijkl}) are arranged such that the derivatives are
manifestly correct in one dimension. For the Kepler problem, where
\be
V(r)=-\frac1r,
\ee
the error Hamiltonians up to the fourth order are:
\ba
H_{TTV}&=&r^{-3}(\delta_{ij}-3\hat\br_i\hat\br_j)p_ip_j\, ,\la{httv}\\
H_{VTV}&=&-r^{-4},\la{hvtv}\\
H_{TTVTV}&=&4r^{-6}(\delta_{ij}-6\hat\br_i\hat\br_j)p_ip_j\, ,\la{httv2}\\
H_{VTVTV}&=&-4r^{-7},\la{hvtv2}\\
H_{TTTTV}&=&-9r^{-5}(\delta_{ij}\delta_{kl}-10\,\delta_{ij}\hat\br_k\hat\br_l+
\frac{35}3 \hat\br_i\hat\br_j\hat\br_k\hat\br_l\
)p_ip_jp_kp_l,\la{ht4v}\\
H_{VTTTV}&=&9r^{-6}(\delta_{ij}-3\hat\br_i\hat\br_j)p_ip_j\la{hvt3v}
\ea
Note that $H_{TTV}$, $H_{TTVTV}$, $H_{VTTTV}$ are all quadratic in
$\bp$ characterize by two numbers $n$ and $\alpha$,
\be
h(n,\alpha)=r^{-n}(\delta_{ij}-\alpha\,\hat\br_i\hat\br_j)p_ip_j.
\la{qham}
\ee
The case of $n=\alpha$ will be shown to be specially simple.

\section {Perihelion advances as perturbative errors}

The basic idea of Sivardi\`ere's method\cite{sivar} of determining the
precession of the Kepler orbit via the rotation of the LRL vector
\be {\bf A}={\bf p}\times{\bf L}-\hat{\bf r}, \ee where
$\hat\br=\br/r$, is to extract the time derivative of \bA\, in the
form of \be
%\frac{d\bA}{d t}=
\dot\bA={\bf \Omega}\times\bA,
\lb{preform}
\ee
thereby identifying the precession angular frequency
${\bf \Omega}$, and obtain the precession angle over one
period by integrating
\be
\Delta\theta=\int_0^P \Omega(t)dt,
\ee
where $P$ is the period. For our purpose, we will generalize
Sivardi\`ere's approach to treat arbitrary, but angular-momentum
conserving forces, including non-separable Hamiltonians.

For any Hamiltonian which leaves ${\bf L}$ invariant, 
\be
\dot\bA=\dot{\bf p}\times{\bf
L}+\frac\br{r^3}\times(\br\times\dot\br). 
\ee 
For the Kepler
Hamiltonian, 
\be H_0=\frac12 \bp^2-\frac1{r}, \lb{hkep} 
\ee 
\be
\dot\br=\bp, \quad \dot\bp=-\frac{\br}{r^3}\quad\Rightarrow
\dot\bA=0. 
\ee 
If (\ref{hkep}) is perturbed by a central force of the form 
\be \dot\bp=-\nabla v(r)=f(r)\hat\br,
\ee 
then one has 
\be \dot\bA=-f(r)\bL\times\hat\br. 
\lb{adot}
\ee 
Without lost of generality, we can always assume that the
unperturbed $\bA$ lies along the x-axis such that $\bA=e{\bf i}$,
whose length is the eccentricity $e$ of the orbit. Thus we can
cast (\ref{adot}) in the form (\ref{preform}) with 
\be {\bf
\Omega}=-f(r)\frac{L}{e}\cos(\theta)\,\hat{\bf L}, 
\ee 
and 
\be
\Delta\theta =\frac1{e}\int_0^{2\pi}(-f(r)r^2)\cos(\theta) d\theta,
\lb{vpotn} 
\ee 
where we have used $L=r^2\dot\theta$. If $f(r)$ can
be expanded in inverse powers of $r$ via 
\be -f(r)r^2=\sum_n\ld_n
r^{-n}, \ee 
where $n=0,1,2,$ etc., then by the use of 
\be
\frac1r=\frac1\wp (1+e\cos\theta)\quad {\rm with}\quad
\wp=L^2=a(1-e^2), \ee 
where $a$ is the semi-major axis, one obtains the
closed-form result 
\be \Delta\theta=
\sum_n\frac{\lambda_n}{\wp^n}C_{n}(e), \lb{dtf} 
\ee where we have
defined 
\be C_n(e) =\frac1{e}
\int_0^{2\pi}(1+e\cos\theta)^{n}\cos\theta d\theta. 
\la{cne} 
\ee 
In table 1, we list the required integral $C_n(e)$ up to $n=8$. Notice
that for an inverse-square force, $n=0$ and $\dt=0$. By partial
integration, it is easy to see that 
\be S_n(e)
=\int_0^{2\pi}(1+e\cos\theta)^{n}\sin^2(\theta)d\theta=\frac1{n+1}C_{n+1}(e).
\la{sne} 
\ee 
From this, one can also derive the following recursion
relation for $C_n(e)$: 
\be
(1+\frac1{n+1})C_{n+1}=(2+\frac1n)C_n-(1-e^2)C_{n-1}. \la{recur} 
\ee

For $H_{VTV}$, corresponding to $-f(r)r^2=4r^{-3}$ we have
\be
\Delta\theta_{VTV}=\frac4{\wp^3}C_3(e)=\frac{4\cdot 3\,\pi}{\wp^3}(1+\frac14 e^2).
\la{dtvtv}
\ee
For $H_{VTVTV}$, corresponding to $-f(r)r^2=4\cdot 7r^{-6}$, we have similarly,
\be
\Delta\theta_{VTVTV}=\frac{4\cdot 7}{\wp^6}C_6(e)
=\frac{4\cdot 7\cdot6\,\pi}{\wp^6}(1+\frac52 e^2+\frac58 e^4 ).
\ee

The other perturbing Hamiltonians are not local potentials, but are
non-separable Hamiltonians with angular-momentum conserving
equations-of-motion,
\ba
\dot\bp&=&f(\br,\bp)\hat\br+g(\br,\bp)(\bp\cdot\br)\bp,\nn\\
\dot\br&=&-g(\br,\bp)(\bp\cdot\br)\hat\br+h(\br,\bp)\bp.
\lb{genham}
\ea
In this case, we have
\be
\dot\bA=-f(\br,\bp){\bf L}\times\hat\br
+g(\br,\bp)(\bp\cdot\br)\bp\times{\bf L}
-\frac{h(\br,\bp)}{r^2}{\bf L}\times\hat\br.
\lb{geneq}
\ee
The last and the third term can be treated as discussed above. It is
only necessary to expand $-fr^2$ and $-h$ in inverse powers of $r$
and invoke (\ref{dtf}).
The middle term requires further attention. We rewrite it as
\be
\dot\bA=g(\br,\bp)(\bp\cdot\br)(\bA+\hat\br)
\ee
The first term above has the exact solution
\be
\bA(t)=\exp\Bigl[\int_0^tg(\br,\bp)(\bp\cdot\br)dt\Bigl]\bA(0),
\ee
which induces no rotation on $\bA$ and can be ignored. For
the second term, relative to $\hat\bL\times\hat\br$, $\hat\br$
lags $90^{\circ}$ behind, so that the corresponding $\bO$ is
given by
\be
\bO=g(\br,\bp)(\bp\cdot\br)\frac{1}{e}\cos(\theta-\frac\pi{2})\,\hat\bL,
\ee
with
\be
\Delta\theta=\frac{1}{e}\int_0^P g(\br,\bp)(\bp\cdot\br)\sin(\theta) dt.
\ee
In doing the time integration, one can use the
unperturbed Kepler orbit, with $\bp\cdot\br=r\dot r$ and
\be
\frac{\dot r}{r^2}=\frac{e}{\wp}\sin(\theta) \dot\theta.
\ee
Hence,
\be
\Delta\theta=\frac{1}{\wp}\int_0^{2\pi} g(\br,\bp)r^3\sin^2(\theta) d\theta.
\ee
If $g$ can be expanded in inverse power of $r$ such that
\be
gr^3=\sum_n\rho_n r^{-n},
\ee
then again we have the closed-form result
\be
\Delta\theta_g=\sum_n\frac{\rho_n}{\wp^{n+1}}S_n(e)
=\sum_n\frac{\rho_n}{\wp^{n+1}}\frac{C_{n+1}(e)}{n+1}.
\lb{dtg}
\ee

For the quadratic Hamiltonian $h(n,\alpha)$, we have
equations-of-motion of the form (\ref{geneq}) with
\ba
-fr^2&=&-nr^{-n+1}\bp^2+\alpha (n+2)r^{-n-1}(\bp\cdot\br)^2,\nn\\
gr^3&=&2\alpha r^{-n+1},\nn\\
-h&=&-2r^{-n}.
\la{eqqham}
\ea
The precession angle from $r^3g$ and $-h$ can be read off
directly:
\ba
 \dt_{g}&=&2\alpha\frac{S_{n-1}(e)}{\wp^n}=2\frac{\alpha}n\frac{C_n(e)}{\wp^n},
      \la{delg}\\
 \dt_{h}&=&-2\frac{C_n(e)}{\wp^n}.
       \la{delh}
\ea
These two contributions exactly cancel if $n=\alpha$.

Since the time integration can be done along the unperturbed Kepler orbit,
we can replace
\be
\bp^2=\frac2{r}-\frac1{a},\quad (\bp\cdot\br)^2=\bp^2r^2-L^2
\la{unpert}
\ee
and reduce $-fr^2$ to only a function of $r$
\be
-fr^2=2(\alpha(n+2)-n)r^{-n}-\frac1a(\alpha(n+2)-n)r^{-n+1}-\alpha(n+2)L^2r^{-n-1},
\ee
yielding
\be
\Delta\theta_{f}=\frac1{\wp^n}
\left[ 2(\alpha(n+2)-n)C_n-(\alpha(n+2)-n)(1-e^2)C_{n-1}
-\alpha(n+2)C_{n+1}\right].
\ee
By the use of recursion relation (\ref{recur}), this can be simplified to
\be
\Delta\theta_{f}=\frac1{\wp^n}
\left[ (1-\frac\alpha{n}(n+2))C_n
+(\alpha-n)\frac{(n+2)}{n+1}C_{n+1}\right].
\ee
For $\alpha=n$, we just have
\be
\Delta\theta_{f}=-\frac1{\wp^n}(n+1)C_n(e).
\la{delf}
\ee
Combining results (\ref{delg}), (\ref{delh}) and (\ref{delf}),
for $H_{TTV}$ ($\alpha=n=3$), we have
\be
\dt_{TTV}=-\frac4{\wp^3}C_3(e),
\ee
which is the exact negative of $\dt_{VTV}$.
For $H_{TTVTV}$ ($\alpha=n=6$), we have
\be
\dt_{TTVTV}=\frac{4\cdot(-7)}{\wp^6}C_6(e),
\ee
which is the exact negative of $\dt_{VTVTV}$.

For $H_{VTTTV}$, $n=6$ and $\alpha=3$, we have
\ba
\dt_{VTTTV}&=&9(\dt_f+\dt_g+\dt_h)\nn\\
&=&9\left[  \frac1{\wp^6}\Bigl(-3C_6-\frac{24}7 C_7\Bigr)+\frac{C_6}{\wp^6}-2\frac{C_6}{\wp^6}\right]\nn\\
&=&-\frac{9\cdot 4}{\wp^6}\Bigl[C_6(e)+\frac{6}7 C_7(e)\Bigr]\nn\\
&=&-\frac{9\cdot 4\cdot 12\pi}{\wp^6}
\Bigl(1+\frac{25}8 e^2+\frac54 e^4+\frac5{128} e^6\Bigr)
\ea

For $H_{TTTTV}$, we have
\ba
-fr^2&=&9\cdot 5\, r^{-4}\Bigl[p^4-14p^2(\bp\cdot\hat\br)^2+21(\bp\cdot\hat\br)^4\Bigr],\nn\\
gr^3&=&3\cdot4\cdot5\,r^{-4}\Bigl[7(\bp\cdot\hat\br)^2-3p^2\Bigr],\nn\\
-h&=&9\cdot4\,r^{-5}\Bigl[p^2-5(\bp\cdot\hat\br)^2\Bigr]. \la{eqt4v}
\ea By use of (\ref{unpert}), all can be expressed in terms of $r$,
yielding correspondingly \ba
\dt_f&=&\frac{9\cdot8\cdot5}{\wp^6}\Bigl(4C_6-4(1-e^2)C_5+(1-e^2)^2C_4\Bigr)\nn\\
      &&+\,\frac{9\cdot7\cdot5}{\wp^6}\Bigl(3C_8-8C_7+4(1-e^2)C_6\Bigr),\nn\\
\dt_g&=&\frac{3\cdot4}{\wp^6}\Bigl(\frac{20}3 C_6-4(1-e^2)C_5-5C_7\Bigr),\nn\\
\dt_h&=&\frac{9\cdot4}{\wp^6}\Bigl(-8\,C_6+4(1-e^2)C_5+5\,C_7\Bigr).
\la{eqt4vr}
\ea
The repeated use of the recursion relation (\ref{recur}) to eliminate
all terms except $C_6$ and $C_7$ simplifies the above to
\ba
\dt_f&=&\frac{9\cdot4}{\wp^6}\Bigl(C_6+\frac47C_7\Bigr),\nn\\
\dt_g&=&\frac{3\cdot4}{\wp^6}\Bigl(-2C_6-\frac37C_7\Bigr),\nn\\
\dt_h&=&\frac{3\cdot4}{\wp^6}\Bigl( 2C_6+\frac97C_7\Bigr),
\la{dtt4v}
\ea
finally giving
\ba
\dt_{TTTTV}&=&(\dt_f+\dt_g+\dt_h)\nn\\
&=&\frac{9\cdot 4}{\wp^6}\Bigl[C_6(e)+\frac{6}7 C_7(e)\Bigr],
\la{dtt4vf}
\ea
which is the exact negative of $\dt_{VTTTV}$.

\section {Numerical Verifications}

By monitoring the rotation of the LRL vector of a given
algorithm when solving the Kepler problem, one can directly
check the analytical results of the last section. For this
purpose, it is useful to employ algorithms with only a single
error Hamiltonian. For example, the second order algorithm $I$
\begin{equation}
{\cal T}_I(\ep)=
%  {\rm e}^{ {1\over 6}\ep \hV}
%  {\rm e}^{ {1\over 2}\ep \hT}
%  {\rm e}^{ {2\over 3}\ep \hV}
%  {\rm e}^{ {1\over 2}\ep \hT}
%  {\rm e}^{ {1\over 6}\ep \hV}
  \exp({1\over 6}\ep \hV)
  \exp({1\over 2}\ep \hT)
  \exp({2\over 3}\ep \hV)
  \exp({1\over 2}\ep \hT)
  \exp({1\over 6}\ep \hV)
\label{foura}
\end{equation}
has modified Hamiltonian\cite{chin97} 
\be
H_{A}^{I}=H_0-\frac{\ep^2}{72}H_{VTV}+O(\ep^4). 
\ee 
Algorithm $II$,
obtained by interchanging $\hT \leftrightarrow \hV$, has Hamiltonian
\be 
H_{A}^{II}=H_0+\frac{\ep^2}{72}H_{TTV}+O(\ep^4) . 
\ee 
By running
both algorithms at smaller and smaller $\ep$, and dividing the
rotation angle of the LRL vector after one period by $\ep^2/72$
until convergence is seen, we can directly test the predicted result
(\ref{dtvtv}). For starting values of $\br=(10,0)$ and
$\bp=(0,1/10)$, such that $\wp=L^2=1$ and $e=0.9$, we have
the theoretical result  
\be
\dt_{VTV}=-\dt_{TTV}=45.33318\,. 
\ee 
Algorithm $I$ at $\ep=P/10000$ with double precision gives 
\be \dt_{I}=-45.33157\,. 
\ee 
Algorithm $II$ at the same step size
produces 
\be \dt_{II}=-45.33316\,. 
\ee 
Both are in excellent
agreement with the theoretical value, including the sign. Each
algorithm causes the LRL vector (and hence the orbit) to rotate
differently in time, but at the end of the period, both algorithms
have rotated the LRL vector by the same amount. This is shown
in Fig. 1.

To test $H_{TTTTV}$ and $H_{VTTTV}$, we consider the following
symmetric, fourth-order forward\cite{nosix} algorithm, 
\be \cT= \dots 
\exp(\ep v_0\hV+\ep^3 u_0\hU) 
\exp(\ep t_1\hT) 
\exp(\ep v_1\hV+\ep^3 u_1\hU) 
\exp(\ep t_2\hT) 
\exp(\ep v_2\hV+\ep^3 u_2\hU), 
\ee 
where we have only indicated operators from the center to the 
right and where 
\be 
v_i\hV+\ep^2 u_i\hU 
\ee 
indicates that one should update
the momentum by compute the force from the 
effective potential\cite{chin97,chinchen03} 
\be
v_iV+\ep^2 u_i\{V,\{T,V\}\}=v_iV-\ep^2 u_i(\nabla V)^2. 
\la{veff} 
\ee 
Here, $U=\{V,\{T,V\}\}$ and has nothing to due with the function defined in Section II.
For positive coefficients $\{t_i\}$ and $\{v_i\}$, 
\be 
t_1=\frac3{10},\quad t_2=\frac15,\quad\
v_0=\frac8{27}, \quad v_1=\frac{125}{432},\quad v_2=\frac1{16}, \ee
\be u_0=\frac{3121}{1710720}, \quad u_1=\frac{1145}{2737152},\quad
u_2=\frac{409}{1520640}, 
\ee 
we have algorithm $III$ with
Hamiltonian 
\be
H_A^{III}=H_0+\frac{\ep^4}{207360}H_{VTTTV}+O(\ep^6). \ee 
This forward time-step algorithm with only a single
fourth-order error term can be easily converted to
a sixth-order forward algorithm\cite{nosix} by solving $H_{VTTTV}$ directly
as discuss below.  For a different set of coefficients 
\be t_1=\frac3{10},\quad
t_2=\frac15,\quad\ v_0=\frac2{27}(4\sqrt{3}-3), \quad
v_1=\frac{25}{108}(\sqrt{3}-3),\quad v_2=\frac1{12}(\sqrt{3}-1), 
\ee
\be 
u_0=\frac{1}{98820}(943-461\sqrt{3}), \quad
u_1=\frac{5}{158112}(481-266\sqrt{3}), \quad
u_2=\frac{1}{87840}(617-344\sqrt{3}), 
\ee 
we have algorithm $IV$ with Hamiltonian 
\be
H_A^{IV}=H_0-\frac{\ep^4}{14400}(7-4\sqrt{3})H_{TTTTV}+O(\ep^6). 
\ee
For the same initial condition as before, we have 
\be
\dt_{TTTTV}=-\dt_{VTTTV}=5933.72\,. 
\la{dtt3vn} 
\ee 
For $III$ and
$IV$, we increase $\ep$ to avoid
machine errors. Running both algorithms at $\ep=T/5000$ gives 
\be
\dt_{III}=-5933.77\quad{\rm and}\quad 
\dt_{IV}=-5933.68, 
\ee 
both are in excellent agreement with the predicted value (\ref{dtt3vn}). 
The rotation of the LRL vector in time is given in
Fig.2. Despite the more complicated structure of the fourth-order
Hamiltonians, the resulting rotations of the LRL vector are
very similar to the second order case. The only discernable difference 
is that since the fourth-order Hamiltonians are more singular, 
the LRL vector rotates over a much narrower range near mid period. 

It has been shown in Ref.\cite{nosix} that for positive coefficients,
it is not possible to have both $\tt$ and $\tv$ vanish and hence
not possible to isolate the error Hamiltonian
$H_{TTVTV}$ or $H_{VTVTV}$ by itself. (Using negative coefficients
would entail too many operators with only numerical, rather than
analytical coefficients.) However, since the effects of
$H_{TTTTV}$ and $H_{VTTTV}$ have been verified, one can check
the theoretical results for $H_{TTVTV}$ and $H_{VTVTV}$ in combination
with $H_{TTVTV}$ and $H_{VTVTV}$ in a general fourth-order
algorithm. We will do this in the next section. For future reference,
for the same initial condition, one has
\be
-\dt_{TTVTV}=\dt_{VTVTV}=1812.98\,. 
\ee

For the second and fourth-order algorithms considered in this
section, the error coefficients
$e_{VTV}$, $e_{TTV}$ and $e_{VTTTV}$, $e_{TTTTV}$, are of 
opposite signs, resulting in algorithms which rotate the LRL vector 
in the same direction. This is not accidental, but a basic feature of 
forward symplectic algorithms to be discussed in the next section.

\section {Symplectic corrector algorithms}

A general second-order, time-reversible algorithm has modified
Hamiltonian,
\be
H_A=H_0+\ep^2(e_{TTV}H_{TTV}+e_{VTV}H_{VTV})+O(\ep^4).
\ee
For example, the velocity form of the Verlet algorithm
\begin{equation}
{\cal T}_{VV}(\ep)=
  \exp( {1\over 2}\ep \hV)
  \exp(  \ep \hT)
  \exp(  {1\over 2}\ep \hV)
\label{vv}
\end{equation}
has $e_{TTV}=1/12$ and $e_{VTV}=1/24$. This allows us to immediately
predict that when it is used to solve the Kepler problem, its
precession angle per period, after being divided by $\ep^2$, must be
$\dt_{TTV}/24=-1.8888$. This is illustrated in Fig.3. In order to
eliminate this second order precession error, one must devise
algorithms with $e_{TTV}=e_{VTV}$. This requirement\cite{chincor} is 
the same as for being a second order symplectic corrector\cite{wis96,mcl962}, 
or process\cite{mar96,mar97,blan99} algorithm. 
More generally, a symplectic integrator $\cT$ of order $n$ is a corrector 
kernel algorithm if it is such that the similarity transformed algorithm
$S\cT S^{-1}$ is of order $n+2$, where $S$ is the corrector or processor.
This is possible only for $\cT$ having equal error coefficients\cite{chincor}
for each pair of error terms $\{T,Q\}$ and $\{V,Q\}$. 
When corrector algorithms are applied to the Kepler problem,
the precession error in each order would cancel after each period
and both the energy and the precession error would be periodic in time.

However, it is not easy to satisfy this second-order ``correctablility" 
requirement of 
\be e_{TTV}=e_{VTV}.
\ee
If either $\{t_i\}>0$ or $\{v_i\}>0$, Chin\cite{chincor} and 
Blanes-Casas\cite{blanes05}
have proved that it is {\it not} possible
to have $e_{TTV}=e_{VTV}$. Moreover, a recent theorem\cite{funda} 
has precisely stipulated that for positive factorization coefficients,
$e_{VTV}$ and $e_{TTV}$ must be separated by a finite, calculable gap. 
If $e_{TTV}=0$, then $e_{VTV}<0$ and
if $e_{VTV}=0$, then $e_{TTV}>0$. 
However, it is easy to force $e_{VTV}$ to
equal $e_{VTV}$ if $H_{VTV}=\{V,\{T,V\}\}$ can be directly added to the potential
as done in (\ref{veff}). For example, the Takahashi-Imada (TI)
integrator\cite{ti}, 
\be
\cT_{TI}= \exp\left ({1\over 2}\ep \hT\right ) \exp\left ( \ep \hV +
{1\over {24}}\ep^3[\hV,[\hT,\hV]] \right ) \exp\left ({1\over 2}\ep
\hT\right ), \label{ti} 
\ee 
has $e_{TTV}=e_{VTV}=-1/24=-0.0416667$.
Its LRL rotation angle in solving the Kepler problem is shown in
Fig.3. The precession error, like that of the energy error, now
returns to zero. If $\{t_i,v_i\}$ are allowed to be negative, then
the following corrector algorithm can also be used, 
\be \cT_{NF}=
\dots \exp(\ep v_0\hV) \exp(\ep t_1\hT) \exp(\ep v_1\hV) \exp(\ep
t_2\hT), \la{nf} 
\ee with 
\be v_0=\frac1{2-2^{1/3}},\quad
t_2=\frac12 v_0,\quad t_1=\frac12-t_2,\quad v_1=t_1, 
\ee 
and
$e_{TTV}=e_{VTV}=-0.0470817$. Its precession error is also shown in
Fig.3, denoted as the non-forward (NF) algorithm. Since its error
coefficients are very close to that of TI, its behavior is also very
similar. Note that this non-forward algorithm requires three force
evaluations (the minimum necessary) which is not very efficient. For
three force evaluations, one can have a fourth-order algorithm
without any second-order errors.  
Omelyan\cite{ome06} has recently shown that the force gradient 
in general can be extrapolated with only one additional force evaluation. Thus
the effort in computing the force gradient is the same as the force. 
We conclude from this discussion
that the TI integrator is likely the optimal second-order algorithm for
integrating Keplerian orbits with two force evaluations.

For a fourth-order time-reversible algorithm, the modified 
Hamiltonian is
\bea
\tH = H_0 &+& \ep^4\,
(
\,\tt\,H_{TTTTV} + \tv\,H_{VTTTV}\, \nn \\
&&\quad +\,\vt\,H_{TTVTV}\,
+\,\vv\, H_{VTVTV}\,
)+O(\ep^6).
\la{hc4th}
\eea
By knowing the error coefficients $\tt, \tv, \vt$ and $\vv$, the
precession error of any fourth-order algorithm can be predicted.
For example, the well known Forest-Ruth algorithm\cite{forest} 
has the same form as (\ref{nf}), but with coefficients
\begin{equation}
t_2={1\over 2}v_1,\quad t_1=\frac12-t_2,
%t_1=-{1\over 2}{{2\,^{1/3}-1}\over{2-2\,^{1/3}}},
\quad v_1={1\over{2-2\,^{1/3}}},
\quad v_0=-2\,^{1/3}v_1 ,
\label{tfr}
\end{equation}
error coefficients
\ba 
&&\tt=-0.00041376, \quad \tv=-0.00868165,\nn\\ 
&&\vt=\ \ 0.00702660,\quad \vv=-0.02604494,
\ea
and precession error
\ba
\dt_{FR}&=&(\tt-\tv)\dt_{TTTTV}+(\vv-\vt)\dt_{VTVTV}   \nn\\
        &=&49.0593-59.9580,\nn\\
		&=&-10.8987,
\ea
which is in good agreement with the observed error\cite{chinkid} 
of -10.8890 computed at $\ep=P/10000$. In contrast, the 
forward algorithm
C\cite{chin97}
\be \cT_C= \dots 
\exp(\ep v_0\hV+\ep^3 u_0\hU) 
\exp(\ep t_1\hT) 
\exp(\ep v_1\hV+\ep^3 u_1\hU) 
\exp(\ep t_2\hT), 
\la{algc}
\ee
where 
\begin{equation}
v_0=\frac14,\quad 
v_1=\frac38,\quad 
u_0=\frac1{192},\quad 
u_1=0,\quad 
t_1=\frac13,\quad 
t_2=\frac16,
\label{tcof}
\end{equation}
has error coefficients
\ba 
&&\tt=-\frac7{51840}=-0.000135, \quad \tv=-\frac1{8640}=-0.000116,\nn\\ 
&&\vt=-\frac7{23040}=-0.000304,\quad \vv=-\frac{11}{46080}=-0.000239,
\ea
and a precession error of only  
\ba
\dt_{C}&=&(\tt-\tv)\dt_{TTTTV}+(\vv-\vt)\dt_{VTVTV}   \nn\\
        &=&-0.114462+0.118033,\la{cancel}\\
		&=&0.003570,
\ea
which is more than three order-of-magnitudes smaller. 
This theoretical value is again in excellent agreement with the 
algorithm's actual error of 0.003565 at $\ep=P/10000$. 
Algorithm C uses only one more force gradient (and therefore only one more force) 
than FR. We have previously demonstrated\cite{chinsante} that algorithm C's precession
error is smaller than recent fourth-order symplectic integrator proposed by
McLachan\cite{mcl95}, Blanes and Moan (recommended in Ref.\cite{mcl02}) 
and Omelyan, Mrylgod and Folk\cite{ome02,ome03}. 

For a fourth-order algorithm, the precession error will return 
exactly to zero only if algorithm is correctable with 
\ba
\tt&=&\tv
\la{corone}\\
\vt&=&\vv.
\la{cortwo}
\ea
This partly explains why algorithm C is so much better than algorithm FR:
its error coefficients are more nearly equal. However, its unusually
small precession error is due also to the near cancellation of two
distinct error types in (\ref{cancel}). 

The equality (\ref{cortwo}) can be 
easily satisfied by redistributing the gradient term in C. For example,
by changing only
\be
u_0=(1-\alpha)\frac1{192}\quad{\rm and}\quad u_1=\frac\alpha{2}\frac1{192},
\ee
with 
\be 
\alpha=\frac9{10},
\la{alval}
\ee 
one changes only 
\be 
\vt=\vv=-\frac1{3840}=-0.000260\,.
\ee
The precession error now goes up to
\ba
\dt_{C\,^\prime}&=&(\tt-\tv)\dt_{TTTTV}   \nn\\
        &=&-0.1144622\,.
\la{tveq}
\ea
While this is in excellent agreement with the observed value of $-0.1144619$ at
$\ep=P/10000$, this is clearly not an improvement over algorithm C. 
Instead of forcing only $\vt=\vv$, one can also choose
\be
\alpha=\frac9{10}-\frac4{15}\frac{\dt_{TTTTV}(e)}{\dt_{VTVTV}(e)}
\ee
so that total precession error vanishes for given initial choice of
the eccentricity $e$. For $e=0.9$, we have
\be
\alpha=0.027225479\, .
\ee
Numerically, the precession error of this tailored algorithm returns to
$\dt=-2.11\times 10^{-6}$ after one period. Since $\alpha=0$ corresponds 
to algorithm C, this algorithm differs only slightly from C. However, the 
slight change is essential for forcing the precession error to zero. Its
precession error is compared to that of C in Fig.4. 

The above tailored algorithm is not a general algorithm because it
requires {\it a priori} knowledge of the eccentricity of the orbit.
For a general corrector algorithm, one must enforce (\ref{corone})
in addition to (\ref{cortwo}).
As in the second order case, the equality (\ref{corone}) 
cannot be satisfied for forward algorithms. One must therefore
keep one of the two error Hamiltonians. We keep the simpler $H_{VTTTV}$
and generalize (\ref{algc}) to
\be \cT_C= \dots 
\exp(\ep v_0\hV+\ep^3 u_0\hU) 
\exp(\ep t_1\hT) 
\exp(\ep v_1\hV+\ep^3 u_1\hU) 
\exp(\ep t_2\hT) 
\exp(\ep^5 w_1\hat W), 
\la{algsc}
\ee
where we have denoted simply, $W=H_{VTTTV}$. 
The coefficient $w_1$ is chosen to satisfy (\ref{corone}).

Since $H_{VTTTV}$ is non-separable, one must solve the general
equation-of-motion (\ref{genham}) implicitly.
However, since this error term is of order $\ep^4$ and 
has a small coefficient $w_1$, any low order scheme with
at most 1 iteration is sufficient. We used a second-order implicit
midpoint scheme\cite{hairer}. (A second order method is needed to
preserve time-reversibility. However, at $\ep=P/10000$, the results 
are unchanged even with no iteration, or with the use of the naive
Euler algorithm.) 
For algorithm C (\ref{tcof}) with (\ref{alval}), we must have 
$w_1=-1/103680$. The resulting precession error indeed returns to
zero, however its error near $t=P/2$ is $\approx 0.1$, which
is unacceptably large. By use of the one-parameter family
of algorithm 4ACB as described in Ref.\cite{chinsante}, we have found the
following, likely optimal, fourth-order symplectic
corrector algorithm 4S,
\begin{equation}
v_0=\frac{23}{48},\quad 
v_1=\frac{25}{96},\quad 
t_1=\frac25,\quad 
t_2=\frac1{10},
\label{topt}
\end{equation}
\begin{equation}
u_0=(1-\alpha)\frac{29}{4608},\quad 
u_1=\frac{\alpha}{2}\frac{29}{4608},\quad 
\alpha=\frac{455}{1102},\quad 
w_1=-\frac1{86400}.
\label{topt2}
\end{equation}
Its precession error is compared to that of C and C$\,^\prime$ in Fig.4.
Algorithm 4S's precession error returns to $3.1\times 10^{-6}$ after
one period and is never more than $8.9\times 10^{-3}$ at any time.
Its error coefficients are
\ba
\tt=\tv=\frac1{28800}&=&0.0000347,\nn\\
\vt=\vv=\frac{53}{437760}&=&0.0001211\,.
\ea
The algorithm evolves in time perserving the constancy of 
the modified Hamiltonian (\ref{hc4th}),
\be
H_0(t)+\ep^4H_4(t)=H_0(0)+\ep^4H_4(0) +O(\ep^6),
\ee
where $H_4$ is the total fourth order error function.
It can be extracted as
\be
H_4(t)-H_4(0)=\lim_{\ep\rightarrow 0}\frac1{\ep^4}\Bigl(H_0(0)-H_0(t)\Bigr).
\ee
The right-hand-side is plotted in Fig.5. Algorithm 
C$\,^\prime$'s error is slight higher than than of C,
while the maximum error of 4S is approximately three times
smaller than that of C. For a more general class of fourth order
forward or gradient algorithms other then 4ACB, see 
Refs.\cite{ome02,ome03,chin4th}.

\section{Conclusions and directions for future research }

When solving physical problems, symplectic integrators approximate the
original Hamiltonian by a modified Hamiltonian with a well-defined
error structure. For time-reversible integrators, the error Hamiltonians
come in pairs in the form of $\{T,Q_i\}$ and $\{V,Q_i\}$. 
There is a clear separation between the mathematics of the algorithm, 
which fixes the error coefficients $e_{TQ_i}$ and $\e_{VQ_i}$, and the
physics of the problem, which determines the error Hamiltonians
$\{T,Q_i\}$ and $\{V,Q_i\}$. In the past, when symplectic integrators
are studied as numerical methods, only the error coefficients are
analyzed so that they can be set to zero. Here, by a 
well-chosen example, we have shown that the physical effects of
the error Hamiltonians determine how the error coefficients should
be chosen. That is, the underlying physics of the problem determines 
the best algorithm for its own solution. 

For solving celestial mechanics problems dominated by Keplerian orbits,
this work shows that the optimal integrators at each order are symplectic
corrector algorithms. Unfortunately, for forward algorithms
without any unphysical backward intermediate time steps, this cannot
be implemented without including extra error Hamiltonians.
In second order, it is easy to include $H_{VTV}$, which is just a
local potential. In fourth order, $H_{VTTTV}$ is a non-separable 
Hamiltonian too cumbersome to be solved in general. 
One must find ways of including $H_{VTTTV}$ without solving it directly. 

The analytical results for the precession error are useful for verifying 
numerical calculations, however, it is a tedious way of
proving the equality $\dt_{TQ_i}=-\dt_{VQ_i}$. It should
be possible to prove this equality without explicitly evaluating
individual precession angles.

We have shown in Ref.\cite{chinsante} that the phase error in the
harmonic oscillator vanishes when $e_{TQ_i}=e_{VQ_i}$. It was simply
not realized in that context that $H_{TQ_i}$ and $H_{VQ_i}$ are
also generating exactly opposite phase angles. From these two examples,
maybe one can prove that for a general Hamiltonian with periodic
orbits, only symplectic corrector algorithms can yield periodic errors
for both the action and the angle variable.

Finally, this work demonstrated that one must rethink the 
usual practice of minimizing the sum-of-square of the error 
coefficients as a mean of optimizing algorithms. The error 
Hamiltonians are not random; they come in pairs with opposite signs.
The error coefficients should therefore be chosen to be pair-wise equal,
{\it i.e.}, one should seek optimal algorithms within the
class of corrector algorithms.

\begin{acknowledgments}
This work was supported in part, by a National Science Foundation
grant No. DMS-0310580.
\end{acknowledgments}
%%%%%%%%%%%%%%%%%%%%%%%%%%%%%%%%%%%%%%%%%%%%%%%%%%%%%%%%%%%%%%%%%%%%%%%%%%%%%%%%%
%\newpage
%\bigskip
%\bigskip
%\centerline{REFERENCES}
%\vspace{.3 truein}

%%%%%%%%%%%%%%%%%%%%%%%%%%%%%%%%%%%%%%%%%%%%%%%%%%%%%%%%%%%%%%%%
\newpage
\begin{figure}
    \vspace{0.5truein}
    \centerline{\includegraphics[width=0.8\linewidth]{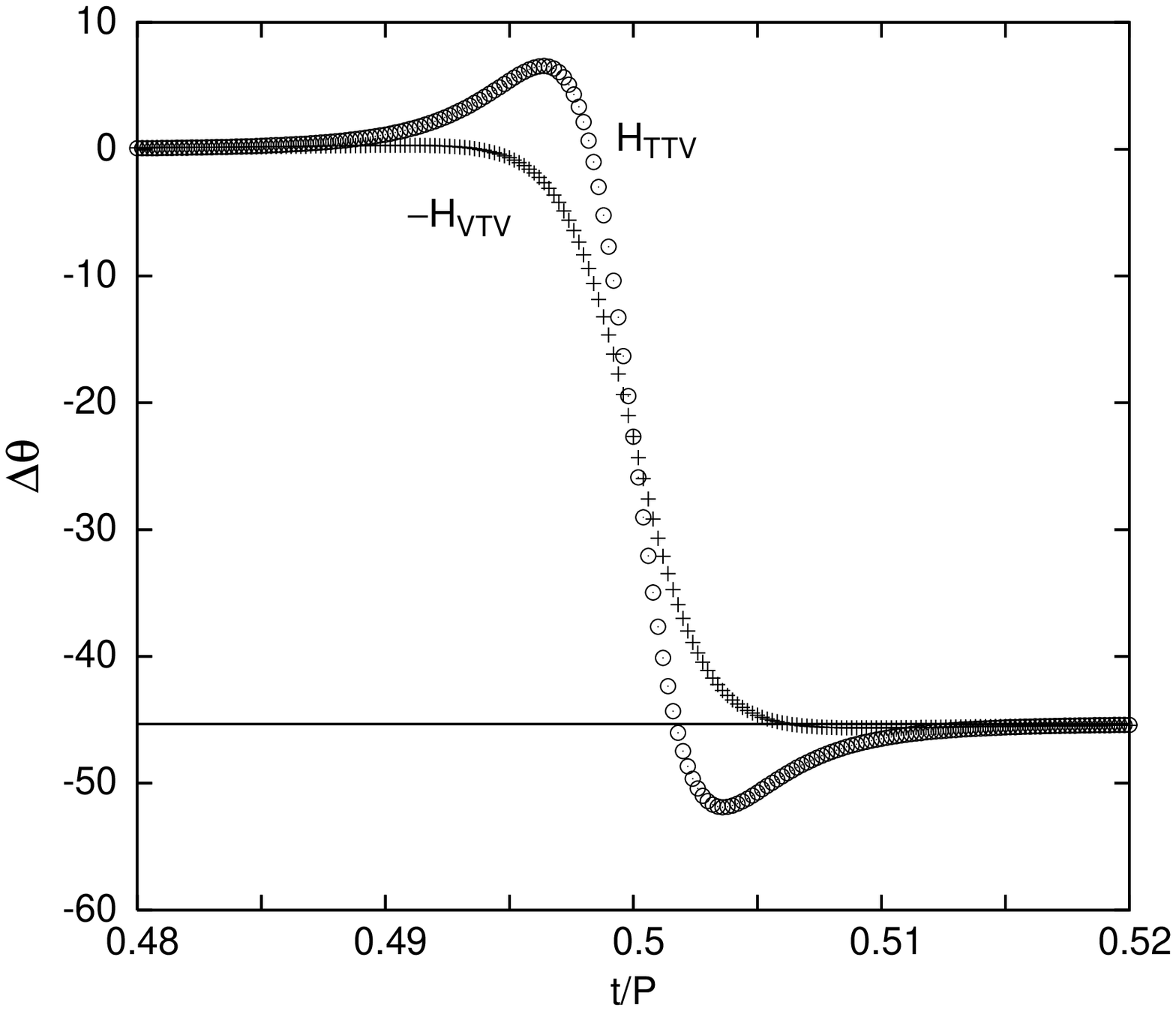}}
    \vspace{0.5truein}
\caption{The rotation of the Laplace-Runge-Lenz vector due to
second order error Hamiltonian
$-H_{VTV}$ and $H_{TTV}$ in algorithms $I$ and $II$. Each algorithm
rotates the LRL vector differently in time, but by exactly the
same amount after one period. Most of the rotation takes
place near the mid period. The solid line gives the
theoretical value of -45.33318.
\label{fig1}}
\end{figure}
%%%%%%%%%%%%%%%%%%%%%%%%%%%%%%%%%%%%%%%%%%%%%%%%%%%%%%%
%%%%%%%%%%%%%%%%%%%%%%%%%%%%%%%%%%%%%%%%%%%%%%%%%%%%%%%%%%%%%%%%
\newpage
\begin{figure}
    \vspace{0.5truein}
    \centerline{\includegraphics[width=0.8\linewidth]{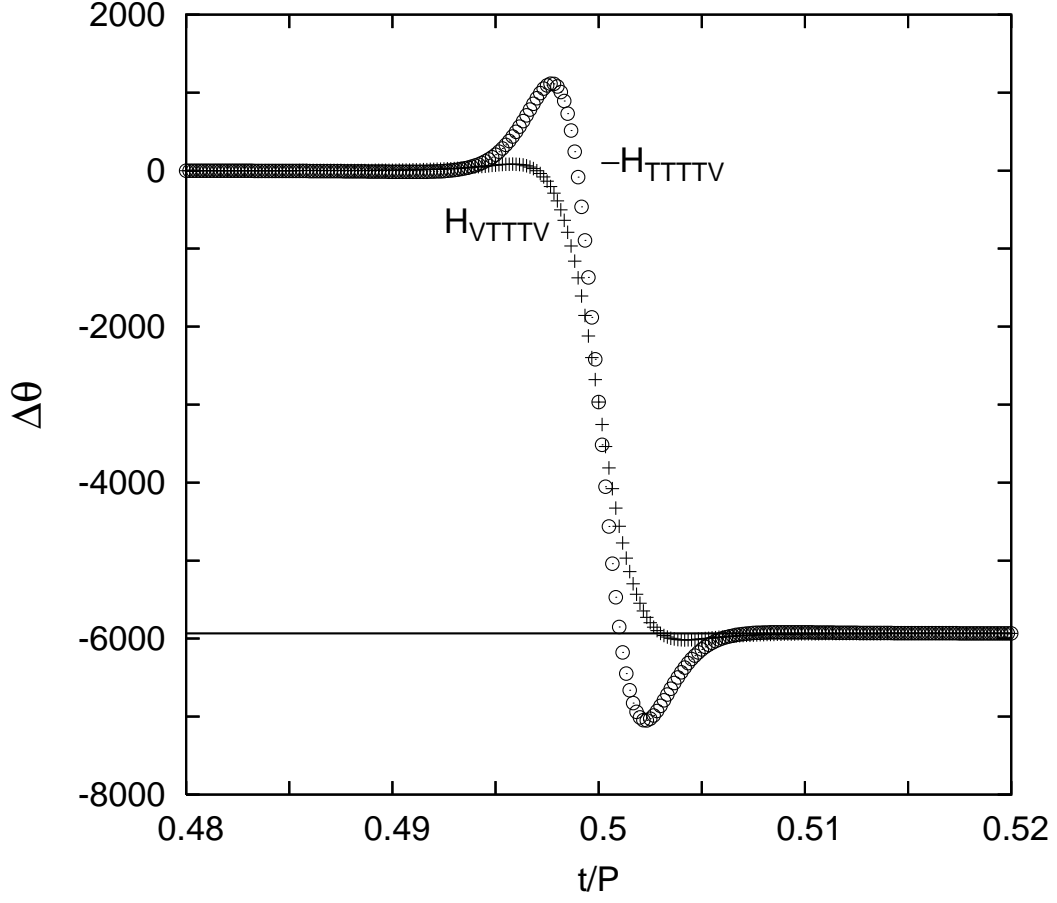}}
    \vspace{0.5truein}
\caption{The rotation of the Laplace-Runge-Lenz vector
due to fourth-order error Hamiltonians
$H_{VTTTV}$ and $-H_{TTTTV}$. Because the fourth-order error terms
are more singular, the rotation takes
place over a narrower range near mid period.
The solid line gives the theoretical value of -5933.72$\,$.
\label{fig2}}
\end{figure}
%%%%%%%%%%%%%%%%%%%%%%%%%%%%%%%%%%%%%%%%%%%%%%%%%%%%%%%%%%%%%%%%
\newpage
\begin{figure}
    \vspace{0.5truein}
    \centerline{\includegraphics[width=0.8\linewidth]{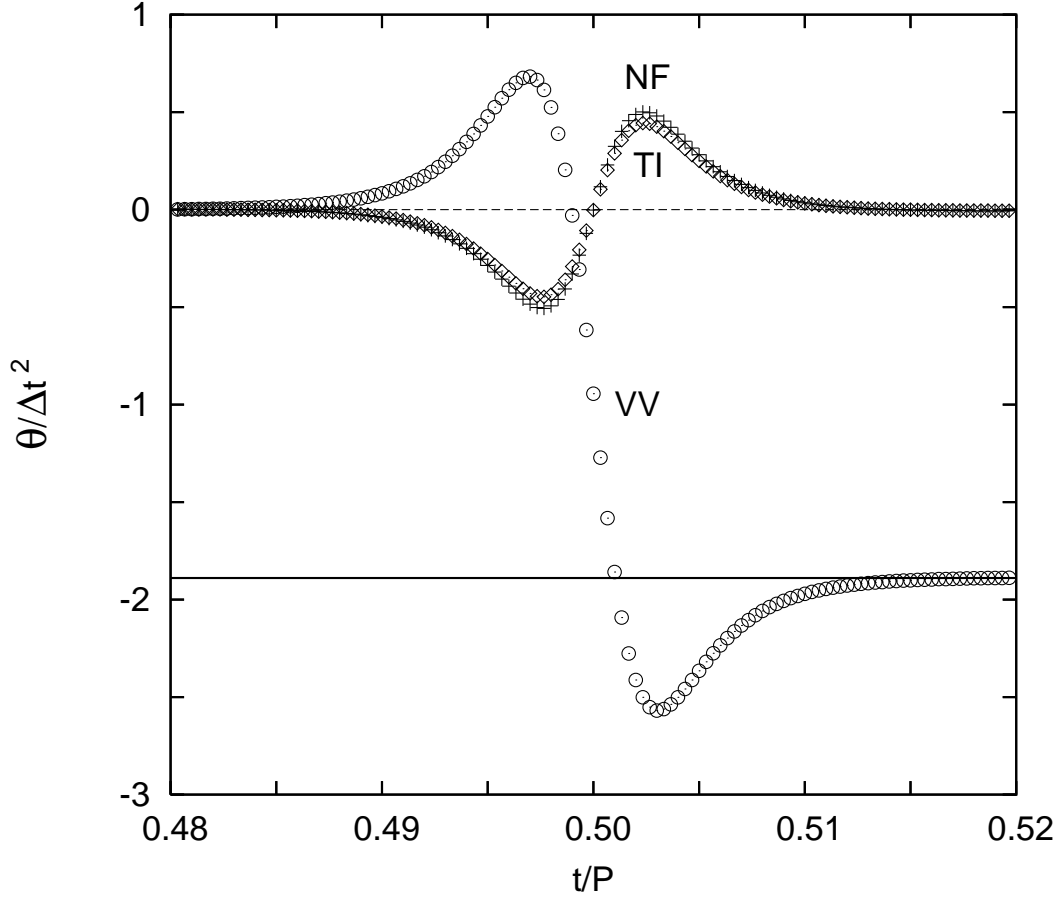}}
    \vspace{0.5truein}
\caption{The rotation of the Laplace-Runge-Lenz vector
for three second-order symplectic algorithms: velocity-Verlet (VV),
Takahashi-Imada (TI) and the non-forward corrrector algorithm (NF).
The solid line gives the theoretical rotation value of the VV algorithm:
$\Delta\theta_{TTV}/24=-1.8888\,$.
\label{fig3}}
\end{figure}
%%%%%%%%%%%%%%%%%%%%%%%%%%%%%%%%%%%%%%%%%%%%%%%%%%%%%%%
%%%%%%%%%%%%%%%%%%%%%%%%%%%%%%%%%%%%%%%%%%%%%%%%%%%%%%%%%%%%%%%%
\newpage
\begin{figure}
    \vspace{0.5truein}
    \centerline{\includegraphics[width=0.8\linewidth]{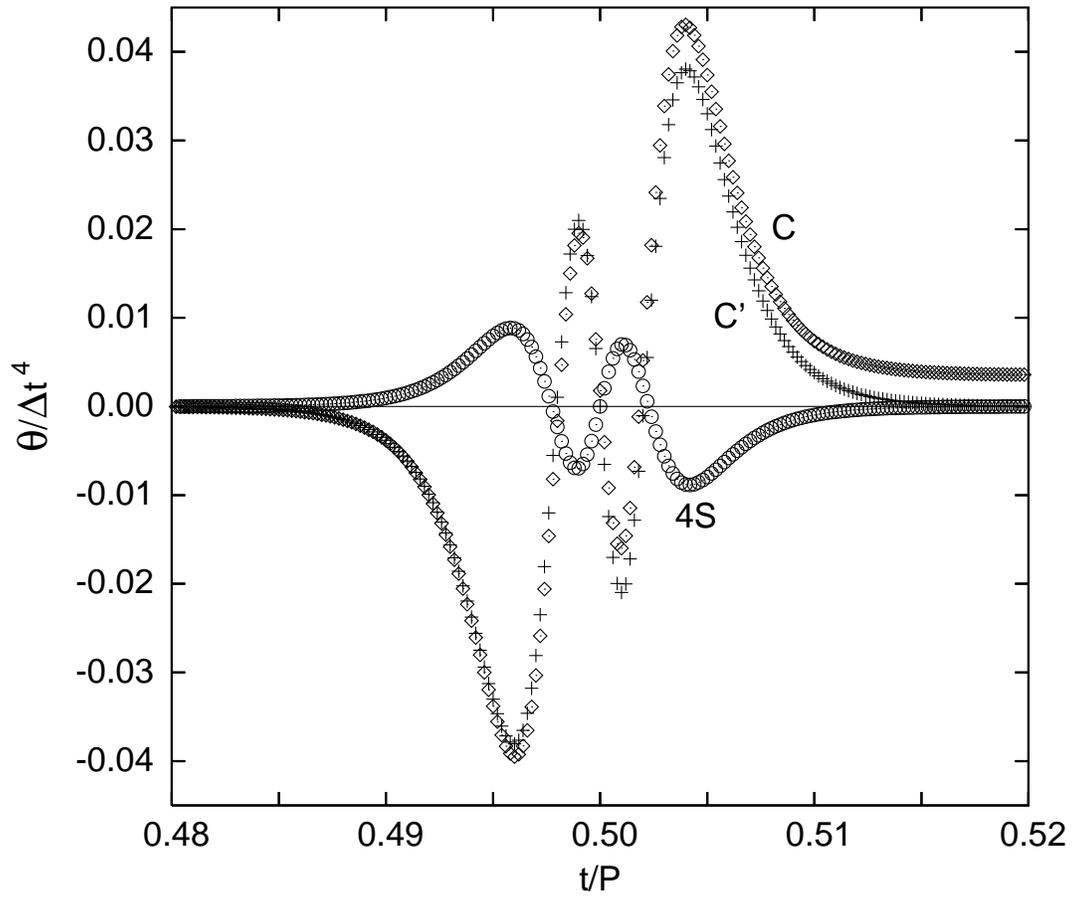}}
    \vspace{0.5truein}
\caption{The rotation of the Laplace-Runge-Lenz vector
for three fourth-order integrators: algorithm C, algorithm C$\,^\prime$
with added gradient term to force the rotation angle back to
zero, and the true symplectic corrector algorithm 4S. 
As with most integrators, algorithm C's
precession error does not return to zero.
\label{fig4}}
\end{figure}
%%%%%%%%%%%%%%%%%%%%%%%%%%%%%%%%%%%%%%%%%%%%%%%%%%%%%%%
%%%%%%%%%%%%%%%%%%%%%%%%%%%%%%%%%%%%%%%%%%%%%%%%%%%%%%%%%%%%%%%%
\newpage
\begin{figure}
    \vspace{0.5truein}
    \centerline{\includegraphics[width=0.8\linewidth]{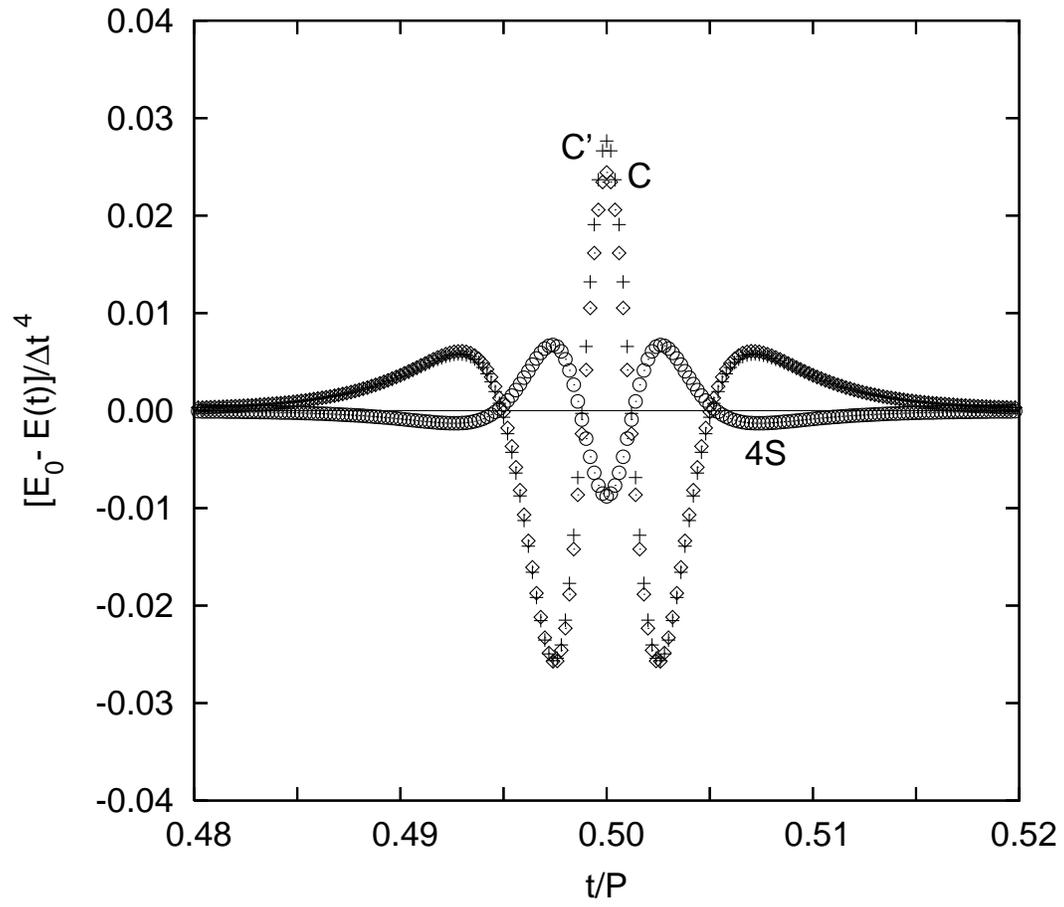}}
    \vspace{0.5truein}
\caption{The energy error functions of algorithms C, C$\,^\prime$
and 4S. Algorithm 4S's maximium error is three times smaller 
than that of C. 
\label{fig5}}
\end{figure}
%%%%%%%%%%%%%%%%%%%%%%%%%%%%%%%%%%%%%%%%%%%%%%%%%%%%%%%
%%%%%%%%%%%%%%%%%%%%%%%%%%%%%%%%%%%%%%%%%%%%%%%%%%%%%%%%%
\begin{table}
\caption{Explicit expressions for the function $C_n(e)$ }
\begin{center}
\begin{tabular}{|c|c|}
\colrule
\colrule
\,\, n\,\, & $C_n(e)$ \\
\colrule
0 & 0 \\
\colrule
1 & $\pi$ \\
\colrule
2 & $2\pi$ \\
\colrule
3 & $3\pi(1+\frac14 e^2 )$\\
\colrule
4 & $4\pi(1+\frac34 e^2 )$\\
\colrule
5 & $5\pi(1+\frac32 e^2+\frac18 e^4 )$ \\
\colrule
6 & $6\pi(1+\frac52 e^2+\frac58 e^4 )$ \\
\colrule
7 &  $7\pi(1+\frac{15}4 e^2+\frac{15}8 e^4 +\frac5{64} )$ \\
\colrule
8 & \quad $8\pi(1+\frac{21}4 e^2+\frac{35}8 e^4 +\frac{35}{64} )$\quad\, \\
\colrule
\colrule
\end{tabular}
\end{center}
\label{tab1}
\end{table}

\end{document}